\documentclass[aps,pra,twocolumn,showpacs,superscriptaddress,groupedaddress]{revtex4-2}
\usepackage{amsmath,bm}
\usepackage{graphicx,epsfig,braket,amssymb,amsfonts,eufrak}

\begin{document}

\title{  Exotic Pairing Structures in Population-Imbalanced Fermionic Systems:  Dynamics as a Probe }
\author{Raka Dasgupta}\email{dasguptaraka@gmail.com}
\affiliation{Dept. of Physics, University of Calcutta, 92 A. P. C. Road, Kolkata-700009, India} 
\author{J.K. Bhattacharjee}\email { jayanta.bhattacharjee@gmail.com}
\affiliation{Indian Association for the Cultivation of Science, Jadavpur, Kolkata-700032, India} 
\begin{abstract}

We investigate a population-imbalanced two-species fermionic system where the resonantly-paired fermions combine to form bosonic molecules via Feshbach interaction. We study the dynamics of the intrinsic quantum fluctuations of the system. It is shown that the natural fluctuations of the condensate fraction consists of a fixed number of periodic components : indicating that these oscillations do not die out, and are sustained in the mean field dynamics of the system. These frequency components bear  distinct signatures of the nature of pairing present in the system. We describe how a time dependent external magnetic field can be used to locate these oscillation frequencies, and thus to explore the momentum space structure of the population imbalanced system. We propose that this method can be used as an indirect experimental probe for detecting exotic phases like the breached pair state, FFLO state, and a phase-separated state comprising of BCS and normal regions. 
\end{abstract}


\maketitle

\section{ INTRODUCTION}

Ultracold two-component Fermi gases have attracted a lot of attention in the last two decades. Feshbach-coupled fermionic systems enjoy an enormous tunability in terms of the interactomic interaction, and can exhibit a crossover from BCS pairing to Bose Einstein condensation(BEC) of fermion pairs \cite{grein,zw1, pat, leggett,eagl, noz,rand2,melo,rann,levin1,ohashi, ohashi2,holland, fuchs, manini,strinati1, sheehy, sheehy2,shin5, raka2}. An interesting variation in this situation is the introduction of an imbalance in the population of the two species. Pairing now has to work around the fact that  not all fermions of type A(one of the species) have a fermion of type B (the other species) to pair with. These can either be two different types of fermionic atoms, or, two hyperfine states of the same atom (for simplicity, we use the $\uparrow$ and $\downarrow$ symbols to denote them): and one has a larger population than the other. This is equivalent to considering superconductivity in the presence of an external magnetic field which creates an imbalance of spin up and spin down states. The consequence can be exotic pairing states like  phase separation \cite{sheehy, sheehy2, shin5}, Breached-Pair (BP) or Sarma phase\cite{liu, bed, wu1, cald,raka}, FFLO \cite{FF,LO, mizu} . 

In the past, the phase diagram of such a spin-polarized Fermi gas has been obtained \cite{shin5} using phase-contrast imaging techniques and a phase-separated state has been detected. However, theoretically predicted exotic phases like FFLO or Breached-Pair have not been observed yet. It is notoriously difficult to detect these phases, characterized by their rich and delicate momentum-space structures.

Several experimental techniques have been proposed previously to identify these phases,  including the study of spatial noise correlations  (to detect BCS states \cite{alt}, for FFLO state in 1 dimensional system \cite{lus}), measurement of occupancy  (to detect FFLO \cite{anna}), momentum-resolved stimulated Raman technique (to detect BP phase       \cite{yi1}), expansion dynamics (to detect FFLO \cite{kaj}). However, there exists no single experimental scheme that can be used to distinguish all the competing phases at once. One such  method has been  proposed by Zhou et al. \cite{zhou} that can differentiate between normal, BCS and BP phases, involving  electromagnetically induced transparency.

In the present paper, we work towards a similar aim, but from a different perspective : we study the fluctuation dynamics of the condensate  and find that it shows an oscillation with a certain number of periodic components. We observe that the structures of the exotic phases introduce new momentum scales in the dynamics, and that, in turn, get reflected in this frequency space.  Thus the oscillation frequencies can be used as a measure to detect phases like FFLO, BP and also the phase-separated state. Moreover, for a specific phase, it can map out the exact details of the momentum-space structure.

In the recent past,  dynamics of ultracold atomic systems remained one of the prime focus of experimental pursuits \cite{grein2, trot, hung, ties, hu1}.  In the theoretical front, oscillatory dynamics in ultracold systems have been studied extensively in different contexts, like the collective nonlinear evolution of a BCS state after an abrupt switching on of the pairing interaction \cite{barankov}, BEC-to-BCS oscillation as the position of the Feshbach resonance is changed abruptly \cite{andreev}, a damped oscillation when Feshbach magnetic field jumps suddenly \cite{burnett}, oscillations in the order parameter as the system is quenched across a quantum critical point \cite{krish}. All these works concentrate on  the response of the system following a rapid change in the system parameter. This entails the study of the higher order nonlinear terms in the evolution equation. We, on the other hand, focus on the natural fluctuations of BEC condensate that occurs along the BCS-BEC crossover path. So we can safely confine ourselves in the linear regime.

The paper is organized as follows. In Sec. II, the basic theoretical model is explained. In Sec. III, the fluctuation dynamics of the condensate corresponding to four distinct representative phases is discussed, and we show how the frequencies of oscillation can be taken as signatures of those phases. In Sec. IV, the origin of these oscillation frequencies are analytically investigated. In Sec V, we propose that a time dependent Feshbach coupling can be used to probe the pairing structures more effectively. The results are summarized in Sec. VI.

\section{MODEL HAMILTONIAN AND EQUILIBRIUM DYNAMICS}

Here we start with  a two-species fermionic system. In addition to the fermion-fermion interaction ( denoted by $g_1$), there is  an additional  interaction ($g_2$) of the Feshbach variety which couples  two fermions to form a bosonic molecule . Our model resembles the one used in \cite{ohashi, holland, raka} .

The Hamiltonian is :

\begin{equation}
\begin{split}
H=\sum_\textbf{p}\epsilon_\textbf{p}({a_\textbf{p}}^\dagger_\uparrow {a_\textbf{p}}_\uparrow+{a_{-\textbf{p}}}^\dagger_\downarrow{a_{-\textbf{p}}}_\downarrow)\\
+g_1\sum_{\textbf{p},\textbf{p}',\textbf{q}}{a_\textbf{p}}^\dagger_\uparrow{a_{-\textbf{p}+\textbf{q}}^\dagger}_\downarrow{a_{-\textbf{p}'+\textbf{q}}}_\downarrow {a_{\textbf{p}'}}_\uparrow \\
+ g_2\sum_{\textbf{p}, \textbf{q}}(b_\textbf{q}^\dagger{a_\textbf{p}}_\uparrow {a_{-\textbf{p}+\textbf{q}}}_\downarrow +{a_\textbf{p}}^\dagger_\uparrow {a_{-\textbf{p}+\textbf{q}}^\dagger}_\downarrow b_\textbf{q})+\epsilon_b \sum_\textbf{q} b_\textbf{q}^\dagger b_\textbf{q}
\end{split}               
\end{equation}

Here $a$ denotes the annihilation operator for the fermions ( ${a_\textbf{p}}_\uparrow$ and ${a_{-\textbf{p}}}_\downarrow$ correspond to two fermionic species), and $b$ represents the bosonic field.  Also, $\epsilon_p=p^2/2m-\mu_F$ and $\epsilon_b=2\nu-\mu_B$,  $2\nu$ being the Feshbach detuning, an experimentally controllable parameter. For a population-balanced system, $\mu_B=2\mu_F$, where $\mu_B$ is the chemical potential for bosons and $\mu_F$ is that of the fermions. 

When formation of  only zero-momentum bosons are considered, we have $\textbf{q}=0$. For $\textbf{q}\neq 0$ states like FFLO, it is possible to describe the system in terms of a single $\textbf{q}$ only. Either way, the summation over $\textbf{q}$ gets dropped. 
 
When both the pairing states are occupied (e.g, in region of BCS pairing), the mean-field equations of motion are :

\begin{equation}
\begin{split}
i\hbar\dfrac{\partial {a_\textbf{p}}_\uparrow}{\partial t}=
\epsilon_p{a_\textbf{p}}_\uparrow
 + g_1\sum_{\textbf{p}'}\langle {a_{-{\textbf{p}'+\textbf{q}}}}_\downarrow {a_{\textbf{p}'}}_\uparrow \rangle {{a_{-\textbf{p}+\textbf{q}}}^\dagger_\downarrow} \\
 +  g_2{{a_{-\textbf{p}+\textbf{q}}}^\dagger_\downarrow} b_\textbf{q}
 \end{split}
 \end{equation}

\begin{equation}
\begin{split}
i\hbar\dfrac{\partial{{a_{-\textbf{p}+\textbf{q}}}_\downarrow}}{\partial t}
=\epsilon_{\textbf{p}-\textbf{q}}{{a_{-\textbf{p}+\textbf{q}}}_\downarrow}  - g_1\sum_{\textbf{p}'}\langle {{a_{-{\textbf{p}'+\textbf{q}}}}_\downarrow} {{a_{\textbf{p}'}}_\uparrow}\rangle {a_\textbf{p}}^\dagger_\uparrow\\
 - g_2{a_\textbf{p}}^\dagger_\uparrow b_\textbf{q}
\end{split}
\end{equation}

Let us now define the expectation value of the pair wavefunction.
\begin{equation}
O_{\textbf{\textbf{p}},\textbf{\textbf{q}}}= \langle{a_{-\textbf{\textbf{p}}+\textbf{q}}}_\downarrow {a_\textbf{p}}_\uparrow\rangle
\end{equation}

Also, the operator $b_\textbf{q}$ is replaced by its expectation value in the ground state, and from this point onwards we denote this expectation value by $b_\textbf{q}$. It turns out that

\begin{equation}
\label{o}
i\hbar\dfrac{\partial O_{\textbf{p},\textbf{q}}}{\partial t}=(\epsilon_\textbf{p} +\epsilon_{\textbf{p}-\textbf{q}})O_{\textbf{p},\textbf{q}} -g_1\sum_{\textbf{p}'} O_{\textbf{p}',\textbf{q}} - g_2b_\textbf{q}
\end{equation}

As for the evolution of $b_q$
\begin{equation}
\label{phi}
i\hbar\dfrac{\partial b_\textbf{q}}{\partial t}=g_2 \sum_{\textbf{p}'} O_{\textbf{p}',\textbf{q}} + \epsilon_b b_\textbf{q}
\end{equation}

A direct analogy with the standard BCS theory \cite{fw} leads to the expression of the gap 
\begin{equation}
\label{geff5}
\Delta_q=g_1\sum_{\textbf{p}'} \langle {a_{-\textbf{p}'+\textbf{q}}}_\downarrow {a_{\textbf{p}'}}_\uparrow\rangle +g_2b_\textbf{q} = g_{\mbox{\scriptsize{eff}}} \sum_{\textbf{p}'}\langle {a_{-\textbf{p}'+\textbf{q}}}_\downarrow {a_{\textbf{p}'}}_\uparrow\rangle  
\end{equation}

$g_{\mbox{\scriptsize{eff}}}$ being the effective pairing interaction. In the static case, let the condensate order parameter be a constant, i.e., if $b_{q0}$ is the equilibrium value of the bosonic field with a particular $q$ value, then $\partial (b_{q0}^\dagger b_{q0})/\partial t=0$.  From Equations (\ref{phi}) and (\ref{geff5}), $g_{\mbox{\scriptsize{eff}}}=g_1-g_2^2/\epsilon_b$. If the fermion-fermion four point interaction is an attractive one (which is indeed the case for a BCS-like superfluid), then taking the effective interection $g_{\mbox{\scriptsize{eff}}}$  in the attractive sense and writing $g_1=-|g_1|$, we arrive at

\begin{equation}
\label{geff}
g_{\mbox{\scriptsize{eff}}}=g_1+\dfrac{g_2^2}{\epsilon_b}
\end{equation}

This expression matches with the standard one that is obtained using diagrammatic methods \cite{ohashi} and variational technique \cite{raka}. Only, here we do not have to use a trial form of the ground state of the system, so the treatment is certainly more general in nature.

\section{FLUCTUATION DYNAMICS : FREQUENCIES OF OSCILLATION }

Next we consider the inherent quantum fluctuations of the system on the top of the base states. Since Eq. (\ref{o}) and Eq. (\ref{phi}) are structurally linear, their counterparts  for the respective fluctuations would be linear as well. Let, $\Tilde{O}_{\textbf{p},\textbf{q}}$ be the fluctuation in $O_{\textbf{p},\textbf{q}}$, and $\Tilde{b}_\textbf{q}$ be the fluctuation in $b_\textbf{q}$. Therefore,

\begin{equation}
\label{o2}
i\hbar\dfrac{\partial \Tilde{O}_{\textbf{p},\textbf{q}}}{\partial t}=(\epsilon_\textbf{p}+\epsilon_{\textbf{p}-\textbf{q}}) \Tilde{O}_{\textbf{p},\textbf{q}} - g_1\sum_{\textbf{p}'} \Tilde{O}_{\textbf{p}',\textbf{q}}-g_2\Tilde{b_\textbf{q}}
\end{equation}
\begin{equation}
\label{phi2}
i\hbar\dfrac{\partial \Tilde{b_\textbf{q}}}{\partial t}=g_2 \sum_{\textbf{p}'} \Tilde{O_{\textbf{p}',\textbf{q}}} + \epsilon_b \Tilde{b_\textbf{q}}
\end{equation}

We take the respective Fourier transforms and find that $\Tilde{O}_{\textbf{p},\textbf{q}}(\omega)=(g_2 \Tilde{b_\textbf{q}}(\omega)+g_1\sum_{\textbf{p}'} \Tilde{O}_{\textbf{p}',\textbf{q}}(\omega))/(\epsilon_\textbf{p}+\epsilon_{\textbf{p}-\textbf{q}}+\hbar \omega)$

So far, the treatment has been general, and the spin-polarization has not been taken into consideration. When there is a population-imbalance in the system, that will correspond to a few specific geometries. Among these probable exotic phases, we consider four candidates : i) BP1 Breached Pair phase, ii) BP2 Breached Pair phase, iii)FFLO phase iv) A phase-separated state, consisting of a superfluid and a normal Fermi liquid made of the remaining unpaired fermions. The idea is : for each of the exotic pairing states, there are certain regions of paired and unpaired fermions , separable either in real space or momentum space. We solve for Eq. (\ref{o2}) and Eq. (\ref{phi2}) in the paired regions, and investigate the dependance of the oscilation frequencis ($\omega$) on the Feshbach detuning $\epsilon_b$. This is controlled by  the span of the paired region (i.e., the amount of imbalance present ) and also the nature of the pairing therein. The $\omega$ vs. $\epsilon_b$ plots thus are indicative of the pairing structures present in the system.

For each if these exotic phases, we inspect a homogeneous system first, and then extend it to treat a trapped system under Thomas-Fermi LDA apprpoximation. In this entire work, we restrict ourselves to mean-field framework.

The measure for population imbalance is defined as polarization $P=(N_{\uparrow}-N_{\downarrow})/(N_{\uparrow}+N_{\downarrow})$. Here we take $N_{\uparrow}$ to be the population of the majority species, and $N_{\downarrow}$ to be the population of the minority species. 

\subsection{BP1 Breached Pair State}

Breached pair is a non-BCS superfluid phase with $\textbf{q}=0$ pairing. It is homogeneous in the real space, but contains ``breached" regions of unpaired fermions in the momentum space. Depending on how the normal (unpaired) and superfluid (paired) regions are organized in the momentum structure, BP states can be classified as BP1  (with a two-shell structure) and BP2 (with a three-shell structure)\cite{yi1}. BP1 state is found to be stable on the BEC region \cite{raka}. This can be linked with the  $SF_M$ state, predicted to be there in the BEC side of the BCS-BEC crossover \cite{sheehy,sheehy2}, consisting  of closed-channel singlet molecules, with the excess unpaired atoms forming a normal Fermi shell.

\begin{figure}[h]
\includegraphics[scale=.5]{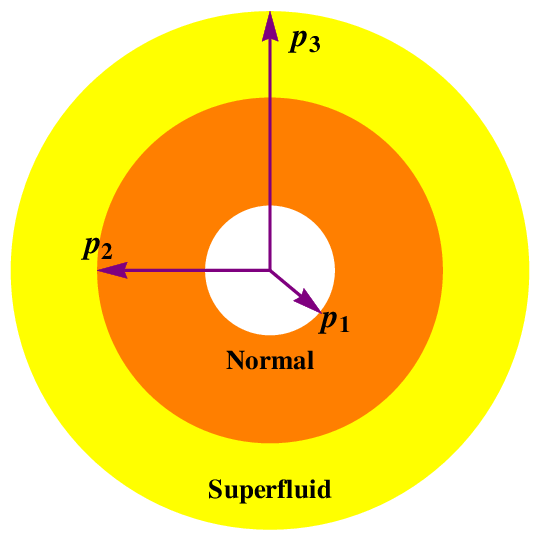}
\caption{Two-shell Structure for Population Imbalanced  BP1 State. Unpaired fermions form the normal region from $p_1$ to $p_2$, and paired fermions form the superfluid region spanning $p_2$ to $p_3$.}
\label{sfmp}
\end{figure}

In Fig. \ref{sfmp}, the shell structure in momentum space is depicted. The unpaired majority fermions stay in the core region, from momenta $p_1$ to $p_2$, which is a normal fluid. The paired superfluid forms the outer shell, between $p_2$ and $p_3$.
Let $p_{F\uparrow}$ and  $p_{F\downarrow}$ to be the Fermi momentum of the majority species and the minority species respectively. However, in the specific structure described above, the highest momentum occupied by the both the atoms is  $p_F$ = $p_{F\uparrow}$. 

Thus, $\Tilde{O_\textbf{p}}(\omega)$ has to be summed over the superfluid region, i.e, over all $p$ from $p_2$ to $p_3$. Here we removed the subscript $\textbf{q}$ from  $\Tilde{O_\textbf{p,q}}(\omega)$ because this state corresponds to zero-momentum pairing, i.e., $\textbf{q}=0$.

\begin{equation} 
\label{sumo}
\sum_\textbf{p} \Tilde{O_\textbf{p}}(\omega)=\big(g_2\Tilde{b}+g_1\sum_\textbf{p} \tilde{O_\textbf{p}}\big )\dfrac{V}{(2\pi \hbar)^3} \int^{p_3}_{p_2}\dfrac{4\pi p^2 dp}{(2\epsilon_p+\hbar \omega)} \\
\end{equation}
Here $V$ is the quantization volume. We measure all length scales in units of $k_F^{-1}$.  From Eq.(\ref{sumo})
\begin{equation}
\begin{aligned}
\begin{split}
\sum_\textbf{p} \Tilde{O_\textbf{p}}(\omega) = & g_2\Tilde{b}(\omega)\frac{u(p_3,\omega)-u(p_2, \omega)}{1-g_1 (u(p_3, \omega)-u(p_2, \omega))}\\
= & g_2 \Tilde{b}(\omega)f_1(p_2,p_3, \omega)
\end{split}
\end{aligned}
\end{equation}
Here 
\begin{equation}
\label{u1}
u(p, \omega)=\dfrac{1}{2\pi^2}\int\dfrac{p^2 dp}{(2\epsilon_p+\hbar\omega)}
\end{equation}
and
\begin{equation}
f_1(p_2,p_3,\omega )=\frac{u(p_3, \omega)-u(p_2,\omega)}{1-g_1 (u(p_3,\omega)-u(p_2,\omega ))}.
\end{equation}
Putting back in Eq.(\ref{phi2}), we obtain 
\begin{equation}
\Tilde{b}(\omega)[\epsilon_b+\hbar\omega+g_2^2 f_1(p_2,p_3,\omega)]=0
\end{equation}
 
Which means, $\Tilde{b}(\omega)$ is zero if $[\epsilon_b+\hbar\omega+g_2^2f_1(p_1,p_2,\omega )]\ne 0$. Therefore, in the expansion of $\Tilde{b}(t)$, only those $\Tilde{b}(\omega)$s will survive for which
\begin{equation}
\label{omega}
\epsilon_b+\hbar\omega+g_2^2f_1(p_2,p_3,\omega )= 0
\end{equation}

In other words,  $\Tilde{b}(t)=b_1 e^{i\omega_1 t}+b_2 e^{i\omega_2 t}+...$,

where $\omega_1$, $\omega_2$ .. are the solutions of equation (\ref{omega}).

Eq. (\ref{omega})  is a nonlinear equation in $\omega$. To find out whether there exist real values for $\omega$, we solve the equation numerically.  We are only interested in real solutions $\omega=\omega_0$, because that would give us solutions in the form of $b(t)=b e^{i\omega_0 t}$, which denotes oscillation. If, on the other hand, we get imaginery solutions for $\omega$, then the solutions are of the form $b(t)=b e^{\omega_0 t}$ or $b(t)=b e^{-\omega_0 t}$. The first one signifies an exponential growth in $b$ and therefore, does not lead to stable solutions. The second one marks exponential decay, and its effect should be negligible as time increases. 

To solve Equation (\ref{omega}) numerically, we use realistic values of the system parameters, corresponding to $^{40} \mbox{K}$ and  $^{6} \mbox{Li}$ atoms respectively, assuming that the two species represent two hyperfine states of the same atom. As discussed in \cite{holland,yi2,koh} the parameters $g_1$, $g_2$, $\epsilon_b$ can be determined from the scattering data, using $g_1=4\pi\hbar^2 a_{bg}n/m$, $g_2=\sqrt{\mu_{co}\Delta B g_1}$ and $\nu=\mu_{co}(B-B_0)$. Here $a_{bg}$ is the background scattering length, $n$ is the density of the fermionic atoms, $\Delta B$ is the width of the resonance. $\mu_{co}$ is the difference in magnetic moment between two channels. We plug in the following parameters : $a_{bg}=174 a_0$, $\Delta B =7.8 G$, $\mu_{co}=1.68\mu_B$ for $^{40} \mbox{K}$, and $a_{bg} =-1405 a_0$, $\Delta B=-300 G$, $\mu_{co}=2\mu_B$ for $^{6} \mbox{Li}$. 

Ideally, the parameters $g_1$, $g_2$, $\nu$ should be properly renormalized to avoid a possible ultraviolet divergence. However, it has been shown that in ultracold systems, the BCS-type pairing can take place within an energy cut-off $E_c=0.541 E_F$ around the Fermi surface \cite{pethick5}. In our model, we employ this cut-off $E_c$ to get rid of the divergence, and safely use the bare coupling parameters. We would like to emphasize that the value of $E_c$ does not enter our calculation explicitly : we just ensure that in the numerical calculations shown later, parameter values are chosen in such a manner,  that all the fermions (irrespective of whether they are paired or not) belong to the region between $E_c$ and $E_F$. Our results are not directly affected by this particular choice of the value of $E_c$. 

\subsubsection{Uniform System}

We use a typical density of $n=10^{14}$  $\text{cm}^{-3}$ \cite{kett}. We scale all energies by the Fermi energy $E_F$, and all momenta by the Fermi momentum $p_F$ (both correspond to the majority species ). Therefore, in this convention, mass of each  partcle gets fixed at 0.5. $\hbar$ is taken as 1. 

A point to note is that, the $n$ that enters in the expression of $g_1$ and $g_2$ is the number of paired atoms. So, this $n$ is determined by the population of the minority species. On the other hand, by Fermi energy, we mean the highest occupied energy level. So that is determined by the population of the majority atoms. Thus, the numerical values of $g_1$ and $g_2$ scaled by $E_F$ also depend on the relative values of $n_\uparrow$ and $n_\downarrow$, i.e., the amount of imbalance in the system. 

\begin{figure}[h]
\includegraphics[scale=.45]{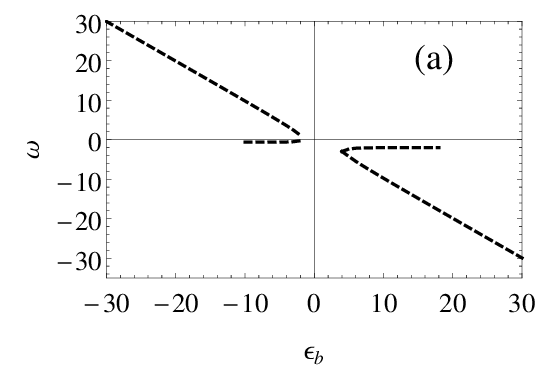}
\includegraphics[scale=.45]{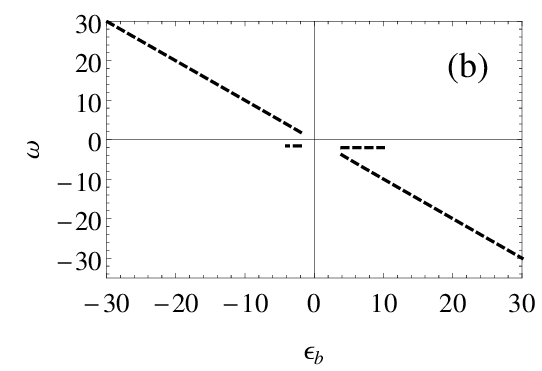}
\includegraphics[scale=.45]{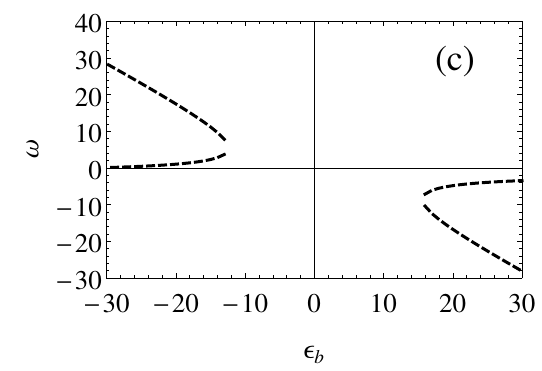}
\includegraphics[scale=.45]{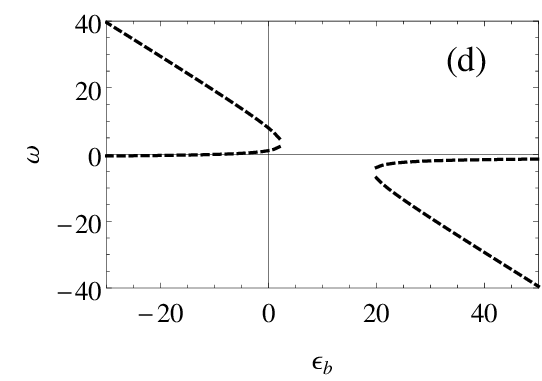}
\caption{$\omega$ vs. $\epsilon_b$ for BP1  state : (a) $^{40} \mbox{K}$, $P=0.3$ (b) $^{40} \mbox{K}$, $P=0.5$ (c) $^{6} \mbox{Li}$, $P=0.3$ (d) $^{6} \mbox{Li}$, $P=0.5$. There are a pair of solutions $\omega_1$ and $\omega_2$ on both sides of the resonance. No real solutions for $\omega$ exist near $\epsilon_b=0$. }
\label{sfm}
\end{figure}

In Fig. \ref{sfm}, the solutions of Equation \ref{omega} are shown as plots of $\omega$ vs.  $\epsilon_b$, corresponding to two atomic systems $^{40} \mbox{K}$ and  $^{6} \mbox{Li}$, each for two polarization values $P=0.3$ and $P=0.5$. We find that for all four cases, there are three regions in the $\omega$ vs.  $\epsilon_b$ plots : 

\begin{enumerate}
\item In the central region near zero detuning, there are no real solutions. The width of this region is smaller for  $^{40} \mbox{K}$ ($\sim$ 0.05 G), and larger for  $^{6} \mbox{Li}$ ($\sim$ 0.2 G).

\item  Away from this region, there are two pairs of allowed $\omega$ values ($\omega_1$ and $\omega_2$), a pair in the positive detuning side, and another in the negative detuning side. Thus the fluctuation in $b(t)$ is $\tilde b(t)=b_1 e^{i\omega_1 t}+b_2 e^{i\omega_2 t}$, or, $b(t)=b_0 + b_1 e^{i\omega_1 t}+b_2 e^{i\omega_2 t}$. So, the condensate fraction (that is given by $|b(t)|^2$) shows an oscillatory dynamics with  frequencies $\pm\omega_1$, $\pm \omega_2$ and  $\pm (\omega_1-\omega_2)$.

\item For a larger magnitude of $\epsilon_b$ (on both detuning sides) only the $\omega_1$ vs. $\epsilon_b$ branch survives, and the $\omega_2\rightarrow 0$. So  $b(t)=b_0 + b_1 e^{i\omega_1 t}$ and the condensate fraction shows an oscillatory dynamics with  frequencies $\pm\omega_1$. This region is shown in the plots for  $^{40} \mbox{K}$. However, in case of $^{6} \mbox{Li}$, this region appears at $\epsilon_b \sim \pm 250$ (that translates to $\pm \sim$ 2.3 G away from the resonance), and is outside the range of the plots shown. 

\end{enumerate} 

\subsubsection{Trapped System}

To treat realistic systems, one would have to include the exact form of the trap potential, which is beyond the scope of our mean-field calculation. However, it can be shown that even in the presence of a trap, the qualitative results remain the same, especially if one focuses on the trap centre.  As in \cite{ohashi2,silva}, the treatment for the bulk, homogeneous system is extended to include a trap using local density approximation (LDA) where the system can be treated to a locally homogeneous one, and the chemical potential $\mu$ is replaced by $\mu(r)=\mu-v_T(r)$, $v_T=(1/2)m\omega_0^2r^2$ being the trap potential. So, Eq. (\ref{sumo})  now has an inbuilt $r$ dependence.
The effective coupling is now given by \cite{ohashi2}
\begin{equation}
g_{\mbox{\scriptsize{eff}}}(r)=g_1+\dfrac{g_2^2}{2\nu-2\mu(r)}
\end{equation} 
 However, the density of the particles is highest at the trap centre ($r=0$) and we focus on that region only. Here, 
\begin{equation}
g_{\mbox{\scriptsize{eff}}}(r=0)=g_1+\dfrac{g_2^2}{\epsilon_b}
\end{equation} 
as in Equation (\ref{geff}).

For the numerical calculation, we consider a simple, homogeneous trap, and assume $\omega=$100 Hz in all three directions. We take a total of $N\sim 10^7$ atoms, a typical value in ultracold experiments \cite{kett}. To correspond with a polarization $P=0.3$, we use $N_\uparrow=657359$ and $N_\downarrow =353963$. This amounts to filling atoms up to $E=(156+3/2)\hbar\omega$, that we call the $E_F$ of the system. All the energies are scaled by this $E_F$.

 The Thomas-Fermi radius is  $R_F=\sqrt{2E_F/m\omega^2}$. Also, the $n_\uparrow$ and $n_\downarrow$ in the homogeneous case are now replaced by $N_\uparrow/R_F^3$ and $N_\downarrow/R_F^3$ respectively, and the parameters $g_1$ and $g_2$ depend on the value of $N_\downarrow$.

\begin{figure}[h]
\includegraphics[scale=.45]{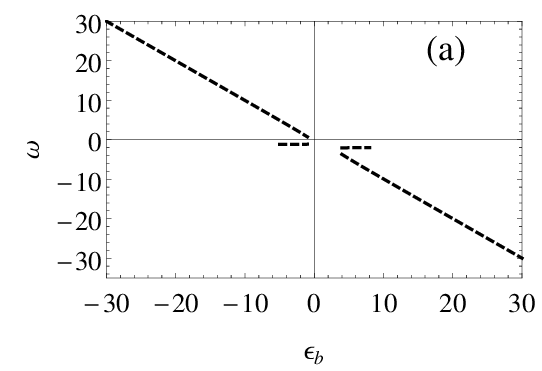}
\includegraphics[scale=.45]{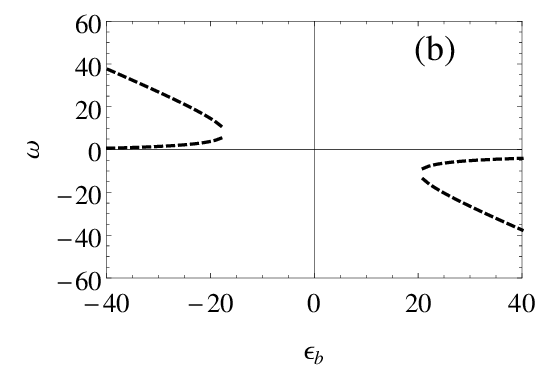}
\caption{$\omega$ vs. $\epsilon_b$ for BP1  state in trap treated under Thomas-Fermi LDA approximation: (a) $^{40} \mbox{K}$, $P=0.3$ (b) $^{6} \mbox{Li}$, $P=0.3$. There are a pair of solutions $\omega_1$ and $\omega_2$ on both sides of the resonance. No real solutions for $\omega$ exist near $\epsilon_b=0$.}
\label{sfmt}
\end{figure}

In Fig. \ref{sfmt},  $\omega$ vs.  $\epsilon_b$ plots are presented for trapped $^{40} \mbox{K}$ and  $^{6} \mbox{Li}$ systems and for  $P=0.3$. Here, too, there are three solution regimes. When $\epsilon_b$ is slightly away from the resonance, the condensate fraction oscillates with  frequencies $\pm\omega_1$, $\pm \omega_2$ and  $\pm (\omega_1-\omega_2)$ . Thus, for both the uniform system and  trapped system, if the  condensate dynamics shows a maximum of three periodic components, BP1 state can be considered to be a strong  possible candidate.  

\subsection{BP2 Breached Pair State}

As mentioned in the previous subsection, the BP2 state is a breached pair state with a three-shell structure, and $q=0$ pairing. In the BCS regime, this state corresponds to the maximum of the thermodynamic potential, and thus, is not a stable one. However, it has  been proposed \cite{son, hu, manna, pao, mishra, guban} that the breached pair state might become stable in the deep BEC regime. Also, in the special case of population-imbalanced p-wave fermionic pairing, BP has been shown to be a possible stable phase along with the usual BCS state \cite {ammar}. 

\begin{figure}[h]
\includegraphics[scale=.5]{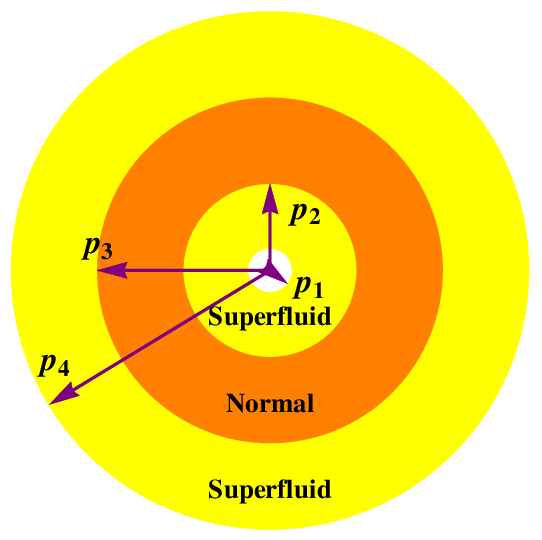}
\caption{ Three-shell structure for Population Imbalanced BP2 State. There is a paired superfluid region from $p_1$ to $p_2$; a breached and unpaired region from $p_2$ to $p_3$; and again another paired superfluid region spanning from $p_3$ to $p_4$.}
\label{bp}
\end{figure}

In Fig. \ref{bp}, the momentum space structure of the breached pair state is shown. There is paired superfluid from $p_1$ to $p_2$, the bairing is breached in the span $p_2$ to $p_3$, and from $p_3$ to $p_4$, there is pairing again.

\subsubsection{Uniform System}

We proceed in the similar way as in the BP1 case, and find that 

\begin{equation}
\epsilon_b+\hbar\omega+g_2^2 f_2(p_1,p_2,p_3,p_4,\omega)=0
\label{omegabp}
\end{equation}
where 

\begin{equation}
\begin{aligned}
\begin{split}
&f_2(p_1,p_2,p_3,p_4,\omega)\\
=&\frac{u(p_4,\omega)-u(p_3,\omega)+u(p_2,\omega)-u(p_1,\omega)}{1-g_1 (u(p_4,\omega)-u(p_3,\omega)+u(p_2,\omega)-u(p_1,\omega))}
\end{split}
\end{aligned}
\end{equation} 

The parameters are same with those in the previous sub-section. Only, now we have a three-shell structure in the momentum space. For the numerical solutions shown, we take $p_1=0.46$ (the lower cut-off) and $p_4=1$ (the Fermi level). We choose $p_2=0.68$, $p_3=0.9$ for polarization $P=0.3$;   and $p_2=0.50$, $p_3=0.9$ for polarization $P=0.5$. 

\begin{figure}[h]
\includegraphics[scale=.45]{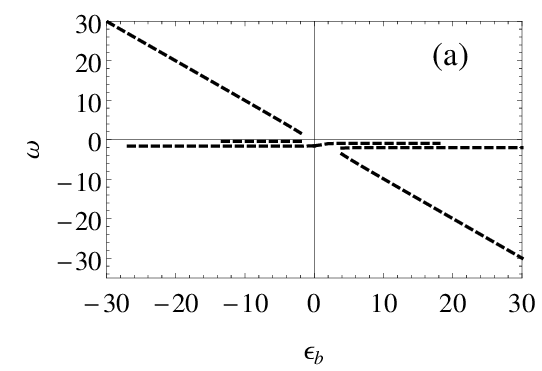}
\includegraphics[scale=.45]{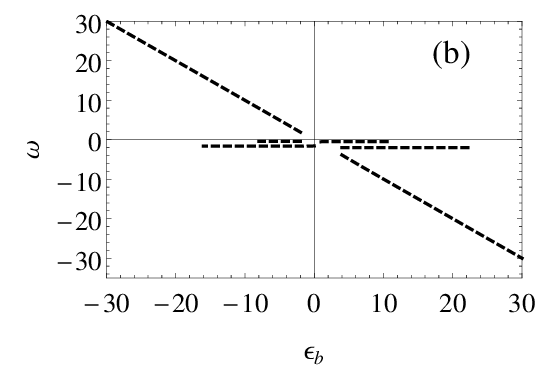}
\includegraphics[scale=.45]{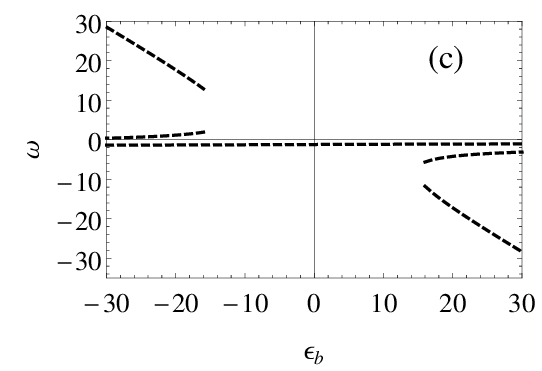}
\includegraphics[scale=.45]{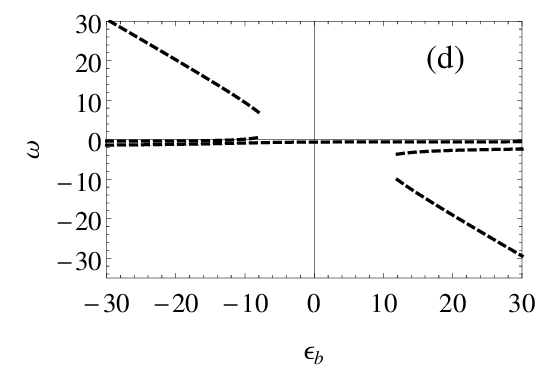}
\caption{$\omega$ vs. $\epsilon_b$ for BP2 state : (a) $^{40} \mbox{K}$, $P=0.3$ (b) $^{40} \mbox{K}$, $P=0.5$ (c) $^{6} \mbox{Li}$, $P=0.3$ (d) $^{6} \mbox{Li}$, $P=0.5$. There are a pair of solutions $\omega_1$ and $\omega_2$ on both detuning sides. These two frequencies are absent near $\epsilon_b=0$. There is also an additional frequency branch of $\omega_3$ spanning the central region including $\epsilon_b=0$ }
\label{bpomega}
\end{figure}

In Fig. \ref{bpomega}, the solutions of Equation \ref{omegabp} are shown as plots of $\omega$ vs.  $\epsilon_b$, corresponding to $^{40} \mbox{K}$ and $^{6} \mbox{Li}$,  and  two polarization values $P=0.3$ and $P=0.5$. We find that for all four cases, there are three solution regimes  :

\begin{enumerate}
\item In the central region near zero detuning, there is one real solutions only. This frequency was not there in the BP1 structure, and we call it $\omega_3$ to distinguish between the $\omega_1$ and $\omega_2$ branches present in BP1.  The width of this single frequency region is smaller for  $^{40} \mbox{K}$ ($\sim$ 0.05 G), and larger for  $^{6} \mbox{Li}$ ($\sim$ 0.2 G). So  $b(t)=b_0 + b_1 e^{i\omega_1 t}$ and the condensate fraction shows an oscillatory dynamics with  frequencies $\pm\omega_1$.

\item  Away from this region, in addition to this $\omega_3$ (which is small in magnitude) there are two pairs of allowed values :  $\omega_1$ and $\omega_2$  for positive and negative detunings both, similar to the BP1 situation.  Thus, $b(t)=b_0 + b_1 e^{i\omega_1 t}+ b_2 e^{i\omega_2 t}+b_3 e^{i\omega_3 t}$. The dynamics of the condensate fraction ($|b(t)|^2$) has six periodic components : $\pm \omega_1$, $\pm \omega_2$, $\pm\omega_3$, $\pm (\omega_1-\omega_2)$, $\pm  (\omega_2-\omega_3)$, $\pm (\omega_3-\omega_1)$.

\item For a larger magnitude of $\epsilon_b$ (on both detuning sides) only one $\omega$ vs. $\epsilon_b$ branch (say, $\omega_1$) survives, as $\omega_2\rightarrow 0$. So  $b(t)=b_0 + b_1 e^{i\omega_1 t}$, and the condensate fraction shows an oscillatory dynamics with  frequencies $\pm\omega_1$. This region is shown in the plots for  $^{40} \mbox{K}$. However, in case of $^{6} \mbox{Li}$, this region appears at $\epsilon_b \sim \pm 250$ (that translates to $\pm \sim$ 2.3 G away from the resonance), and is outside the range of the plots shown. 

\end{enumerate}

We find that there can be a maximum of three branches of allowed $\omega$ values. In addition to the pairs of $\omega$ branches as in the BP1 situation, there appears one additional $\omega_3$ branch at the resonance in this case. Thus, the  dynamics of the condensate fraction ($|b(t)|^2$) has a maximum of six periodic components : $\pm \omega_1$, $\pm \omega_2$, $\pm\omega_3$, $\pm (\omega_1-\omega_2)$, $\pm  (\omega_2-\omega_3)$, $\pm (\omega_3-\omega_1)$. 
We would like to add that although $\omega_3$ is very small and appears to be $\approx 0$, this is not a trivial mode and can be detected separately. An $|\omega_3| \approx 0.5$ (as in Fig. \ref{bpomega}) in our unit (with $\hbar =1$, $m=0.5$ and $E_F=1$) translates to an actual frequency of $\approx$ 60 KHz. So the $\omega_1$ and ($\omega_1-\omega_3$) components (and similarly for $\omega_2$ and ($\omega_2-\omega_3$)) differ by $\approx$ 60 KHz and can certainly be distinguished in experiments. 

\subsubsection{Trapped System}

\begin{figure}[h]
\includegraphics[scale=.45]{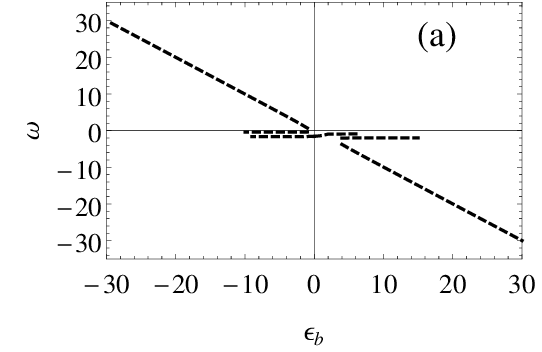}
\includegraphics[scale=.45]{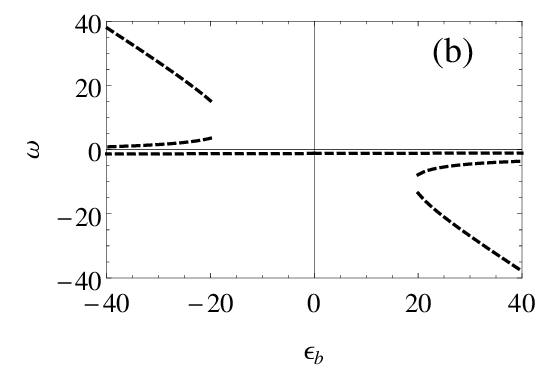}
\caption{$\omega$ vs. $\epsilon_b$ for BP  state in trap treated under Thomas-Fermi LDA approximation: (a) $^{40} \mbox{K}$, $P=0.3$ (b) $^{6} \mbox{Li}$, $P=0.3$. There are a pair of solutions $\omega_1$ and $\omega_2$ on both detuning sides. These two frequencies are absent near $\epsilon_b=0$. There is also an additional frequency branch of $\omega_3$ spanning the central region including $\epsilon_b=0$. }
\label{bpt}
\end{figure}

The same approach (as in BP1) is used here, and the system is treated under Thomas-Fermi LDA method. \\

The $\omega$ vs.  $\epsilon_b$ plots are presented in  Fig. \ref{bpt}, for trapped $^{40} \mbox{K}$ and  $^{6} \mbox{Li}$ systems, using $P=0.3$. Here, too, we find that there is a regime with $b(t)=b_0 + b_1 e^{i\omega_1 t}+b_2 e^{i\omega_2 t}+b_3 e^{i\omega_3 t}$, so, a maximum of six periodic components in the dynamics of $|b(t)|^2$.

So for both the  uniform system and trapped system, a condensate dynamics with six distinct periodic components can be considered to be a signature of the BP2  state. 
 
\subsection{FFLO Phase}
In the Fulde-Ferrell-Larkin-Ovchinnikov (FFLO) state, Cooper pairs have finite and non-zero center-of-mass-momentum. In the Fulde-Ferrell (FF) type of \cite{FF} superfluid, all pairs have the same momentum $\textbf{q}$, and the value of $\textbf{q}$ depends on the amount of imbalance present in the system.  In the Larkin-Ovchinnikov (LO) picture \cite{LO}, the pairing takes place with  momenta $\pm \textbf{q}$ . LO state usually has a lower energy than the FF state and thus, is more stable \cite{Mora, yip, stoof}. However, it is argued that since FF and
LO ansatzes both lead to the same window of existence of the superfluid in the parameter space, the simpler FF picture can be successfully employed to probe various properties of the FFLO state \cite{Jami}. In our present work, we, too use the FF picture with $|\textbf{q}|$ varying as $|\textbf{k}_{F\uparrow}-\textbf{k}_{F\downarrow}|$.

\begin{figure}[h]
\includegraphics[scale=.45]{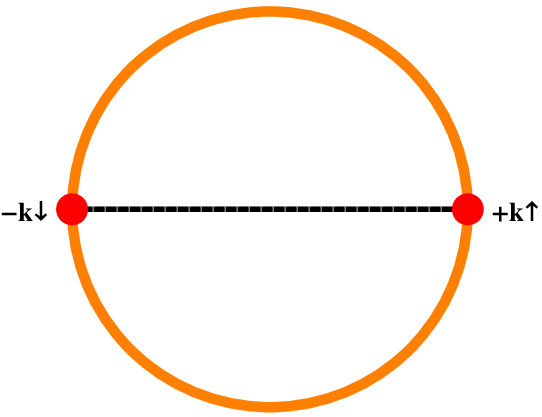}
\includegraphics[scale=.4]{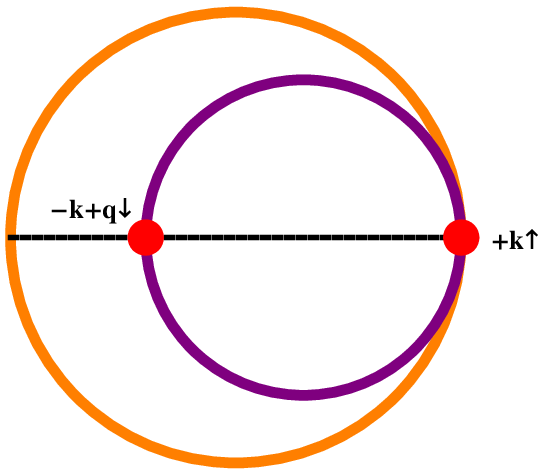}
\caption{The pairing structure in  (a) BCS Superfluid (b) FFLO Superfluid}
\label{ff}
\end{figure}
Here, there is no ``breached" region, and the state supports $q\neq 0$ pairing structure. Unlike the equivalent BCS state where $ \Tilde{O_{p,q}}(\omega)$ has to be summed from $p=0$ to $p=p_F$, for FFLO the sum would be from $p=q$ to $p=p_F$. 

Here the corresponding equations are : 
\begin{equation}
\begin{aligned}
\begin{split}
 \sum_\textbf{p} \Tilde{O_{\textbf{p,q}}}(\omega) =& g_2\Tilde{b}(\omega)\frac{u(p_F,q, \omega )-u(q,q,\omega)}{1-g_1 (u(p_F,q,\omega)-u(q,q,\omega ))}\\
=&g_2 \Tilde{b}(\omega)f_3(p_F,q,\omega)
\end{split}
\end{aligned}
\end{equation}

Here 
\begin{equation}
u(p,q)=\dfrac{1}{2\pi^2}\int\dfrac{p^2 dp}{(\epsilon_\textbf{p}+\epsilon_{\textbf{p-q}}+\hbar\omega)}
\end{equation}
and
\begin{equation}
f_3(p_F,q)=\frac{u(p_F,q)-u(q,q)}{1-g_1 (u(p_F,q,\omega)-u(q,q,\omega))}.
\end{equation}
Putting back in Eq.(\ref{phi2}), we obtain 
\begin{equation}
\Tilde{b}(\omega)[\epsilon_b+\hbar\omega+g_2^2 f_3(p_F,q,\omega )]=0
\end{equation}

\subsubsection{Uniform System}

\begin{figure}[h]
\includegraphics[scale=.45]{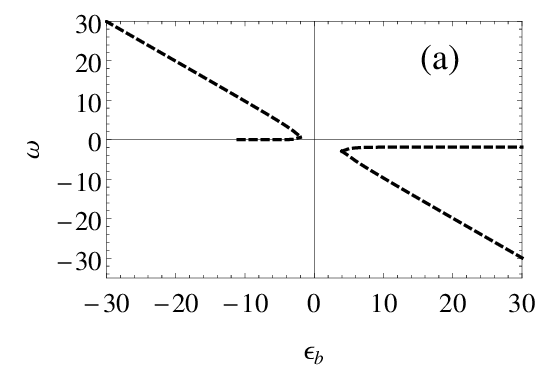}
\includegraphics[scale=.45]{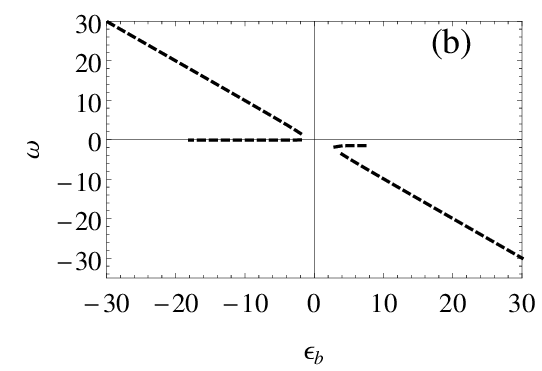}
\includegraphics[scale=.45]{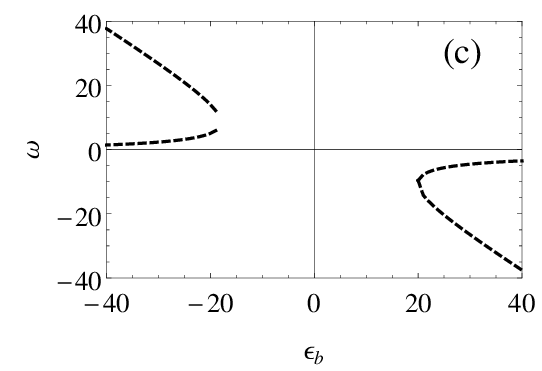}
\includegraphics[scale=.45]{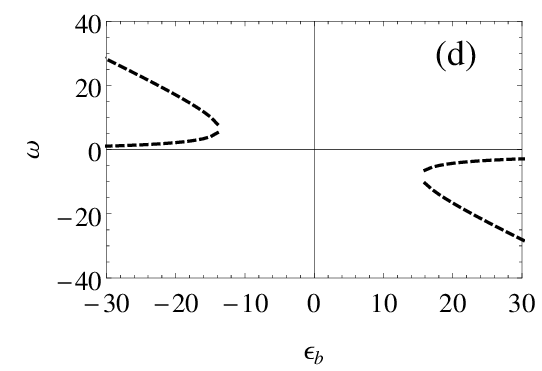}
\caption{$\omega$ vs. $\epsilon_b$ for FFLO state : (a) $^{40} \mbox{K}$, $P=0.3$ (b) $^{40} \mbox{K}$, $P=0.5$ (c) $^{6} \mbox{Li}$, $P=0.3$ (d) $^{6} \mbox{Li}$, $P=0.5$. There are a pair of solutions $\omega_1$ and $\omega_2$ on both sides of the resonance, as in the BP1 state. However, the $\omega$ values are shifted from the BP1 $\omega$ values, and the amount of shift depends on the FFLO momentum $\textbf{q}$. No real solutions for $\omega$ exist near $\epsilon_b=0$.}
\label{fflo}
\end{figure}

The $\omega$ vs. $\epsilon_b$ curves for FFLO in a uniform system are shown in Fig. (\ref{fflo}). These are similar in nature as the plots in the BP1 case , i.e., there are three solution regimes, and the maximum number of periodic components that the system can support is three. Only, for a fixed $\epsilon_b$, the allowed $\omega$ values here get slightly shifted from the $q=0$ BP1 case.

\subsubsection{Trapped System}

\begin{figure}[h]
\includegraphics[scale=.45]{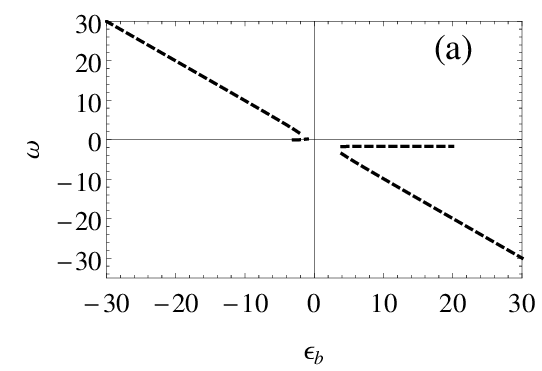}
\includegraphics[scale=.45]{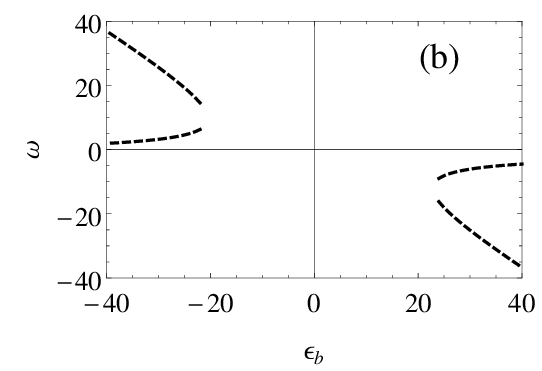}
\caption{$\omega$ vs. $\epsilon_b$ for FFLO  state in trap treated under Thomas-Fermi LDA approximation : (a) $^{40} \mbox{K}$, $P=0.3$ (b) $^{6} \mbox{Li}$, $P=0.3$. There are a pair of solutions $\omega_1$ and $\omega_2$ on both sides of the resonance, as in the BP1 state. However, the $\omega$ values are shifted from the BP1 $\omega$ values, and the amount of shift depends on the FFLO momentum $\textbf{q}$. No real solutions for $\omega$ exist near $\epsilon_b=0$.}
\label{fft}
\end{figure}
The system, treated under Thomas-Fermi LDA approach, shows that at the trap centre, the condensate would oscillate with a pair of $\omega$ values in each side of the resonance. The numerical solutions are plotted in Fig. (\ref{fft}).

\subsubsection{The Effect of Imbalance}
 As mentioned before, the $\omega$ values in the fluctuation of $b(t)$ for the FFLO structure is slightly shifted from the corresponding $\omega$ value for the BP1 (with $q=0$ pairing) state. The amount of this shift depends on the magnitude of $\textbf{q}$, which, in turn, depends on the amount of imbalance present in the system.
 
\begin{figure}[h]
\includegraphics[scale=.45]{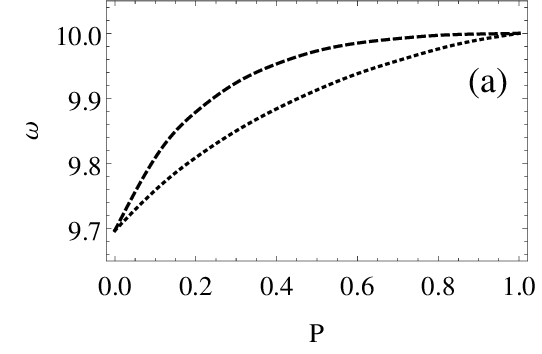}
\includegraphics[scale=.45]{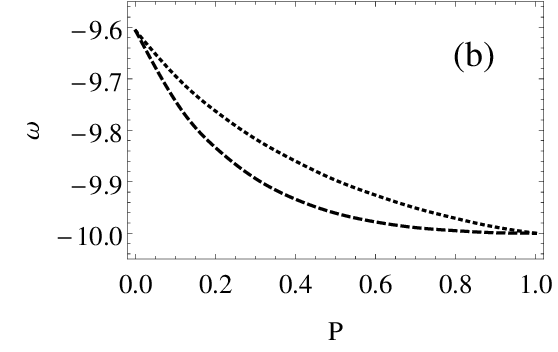}
\caption{Variation of $\omega$ with a varying population imbalance at a fixed $\epsilon_b$,   in a BP1 state (dashed line)  and an FFLO state (dotted line) for a particular branch with $^{40} \mbox{K}$ : (a)  $\epsilon_b=10$ (b) $\epsilon_b=-10$. $P$  varies from 0 (no imbalance) to 1 (full imbalance). }
\label{ffimb}
\end{figure}

  In Fig. (\ref{ffimb}), $\omega$ values are shown for both FFLO and BP1 states on the same plot, with a fixed $\epsilon_b$ (we take $\epsilon_b=10$ and $\epsilon_b=-10$) and a varying polarization $P$. It is found that the two curves meet at $P=0$ and $P=1$. This is because $P=0$ corresponds to the fully paired BCS configuration, and $P=1$ corresponds to a case where only one species is present. In both these limits, BP1 and FFLO pictures lead to the same structure in the momentum space. On the other hand, at $P=0.5$, the deviation between the BP1 solution and the FFLO solution is the largest. 

\subsection{Phase Separated State}
A phase separated state in a population-imbalanced fermionic system  is a combination of a BCS-superfluid, and a free Fermi gas consisting of the imbalanced fermions : separated in the real space.  The phase-separation in an imbalanced gas of $^{6} \mbox{Li}$ was observed by Shin et al \cite{shin5}. They found a shell structure where the superfluid core is surrounded by a normal region. This  state is a consequence of the trap itself, so we do not address the uniform system in this context and directly discuss the geometry of the trapped system.  We consider a superfluid sphere of radius $r_1$ and a  outer spherical shell (between $r_1$ and $r_2$) of normal fermions, completely separated in the real space. Let us assume that the superfluid density is $\rho_s(r)$ and the density of the normal Fermi gas is $\rho_n(r)$.  Now between $r=0$ and $r=r_1$,  $\rho_s(r)= 1$ and $\rho_n(r)=0$ ;  and between $r=r_1$ and $r=r_2$ ,  $\rho_s(r)= 0$ and $\rho_n(r)=1$. Taking a 3-dimensional Fourier transform, we find that
\begin{equation}
\rho_s(p,r_1)= 4\pi\Big(\dfrac{\sin p r}{p^3}-\dfrac{r \cos pr}{p^2}\Big ) {\Bigg |_0^{r_1}}
\end{equation}

Similarly,\\
\begin{equation}
\rho_n(p,r_1,r_2)=4\pi \Big(\dfrac{\sin p r}{p^3}-\dfrac{r \cos pr}{p^2}\Big) {\Bigg |_{r_1}^{r_2}}
\end{equation}

Since the two phases are separated in the real space, we assume that the dynamics of one phase is independent of the dynamics of the other. So when we are talking about the momentum distribution of the superfluid phase, we can effectively treat the normal phase as being non-existent (i.e, $\rho_s(p,r_1)$ is not influenced by $\rho_n(p,r_1,r_2)$).

\begin{equation} 
\label{sumops}
\begin{aligned}
\begin{split}
\sum_\textbf{p} \Tilde{O_\textbf{p}}(\omega)
=&-g_2\Tilde{b}(\omega)\frac{u_4(p_2,r_1,\omega)-u_4(p_1,r_1,\omega)}{1-g_1 (u_4(p_2,r_1,\omega)-u_4(p_1,r_1,\omega))}\\
=&-g_2 \Tilde{b}(\omega)f_4(p_1,p_2,r_1,\omega )
\end{split}
\end{aligned}
\end{equation}
Where the paired region spans from $p_1$ to $p_2$ is the momentum space. 
Here 
\begin{equation}
\label{ups1}
u_4(p,r_1,\omega)=\int\dfrac{\rho_s(p,r_1) p^2 dp}{(2\epsilon_p+\hbar\omega)}
\end{equation}
and 
\begin{equation}
f_4(p_1,p_2,r_1,\omega)=\frac{u_4(p_2,r_1,\omega)-u_4(p_1,r_1,\omega)}{1-g_1 (u_4(p_2,r_1,\omega)-u_4(p_1,r_1,\omega))}.
\end{equation}

Inserting in Eq.(\ref{phi2}), we obtain 
\begin{equation}
\Tilde{b}(\omega)[\epsilon_b+\hbar\omega+g_2^2 f_4(p_1,p_2,r_1,\omega)]=0
\label{omegaps}
\end{equation}

\begin{figure}[h]
\includegraphics[scale=.45]{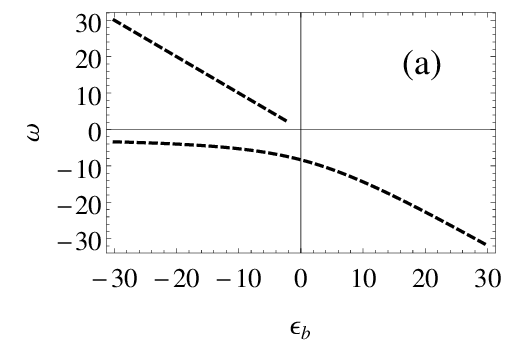}
\includegraphics[scale=.45]{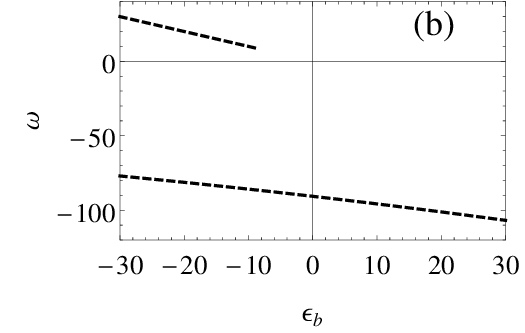}
\caption{$\omega$ vs. $\epsilon_b$ for pase separated  state in trap : (a) $^{40} \mbox{K}$, $P=0.3$ (b) $^{6} \mbox{Li}$, $P=0.3$. There is a single $\omega_1$ branch in the positive detuning side, and two  branches ($\omega_1$ and $\omega_2$) in the negative detuning side.}
\label{plotps}
\end{figure}

Obviously, the solution depends on the value of $r_1$. From \cite{shin5}, we get that  an experiment with $P=0.3$ yields $r_1\sim 35 \mu$m and $r_2\sim 65 \mu$m, and we use those values here. Also, we equate $p_2$ with $p_F=1$ and $p_1$ with the $p_c$. The numerical solutions of Eq. (\ref{omegaps}) for $P=0.3$ are plotted in Fig. \ref{plotps}. 

\begin{enumerate}
\item In the negative detuning side (excluding a small region near the resonance) , there are two real solutions for $\omega$. So, the condensate fraction (that is given by $|b(t)|^2$) shows an oscillatory dynamics with  frequencies $\pm\omega_1$, $\pm \omega_2$ and  $\pm (\omega_1-\omega_2)$.

\item  In the positive detuning side (including the resonance and extending to a small part of the negative detuning side) there is one allowed $\omega$ solution. So the oscillatory dynamics takes place with frequency $\pm\omega_1$.
\end{enumerate}

This is in clear constrast with the BP1, BP2 and FFLO states where the $\omega$ vs. $\epsilon_b$ curves  looked somewhat symmetric on both sides of the resonance. So it is easily possible to distinguish this phase from other competing phases by looking at the frequency curves.

\section{ORIGIN OF THE MULTIPLE BRANCHES OF FREQUENCY}
In Sec. III, we presented the $\omega$ vs. $\epsilon_b$ plots for different pairing phases for certain choices of population imbalances. Evidently, there are certain generic features, applicable to all the pairing states. 

\begin{enumerate}
\item A frequency branch $\omega \approx -\epsilon_b$ is present for all four phases (in a single side of the resonance for the phase separated state, and on both sides for BP1, BP2 and FFLO). 
\item A pair of solutions appear for each detuning value for BP1, FFLO, BP2 (in this case, there is a third possible solution as well) and phase separation (in the negative detuning side only). When the magnitude of this detuning is large, one of these solutions approaches zero, and only one branch for non-zero values of $\omega$ survives. The branch that survives throughout corresponds to the $\omega \approx -\epsilon_b$ solutions mentioned above.
\item Exactly at resonance and in the near vicinity, there are no $\omega$ solutions (except in the 3rd branch that appears only for BP2).
\end{enumerate} 

The fact that the fluctuations contain distinct periodic components indicates that the fluctuations do not die out with time. Instead, these oscillations persist, and are part of the mean-field dynamics. 

\subsection{A comparison among different pairing structures}
The $\omega$ vs. $\epsilon_b$  plots show certain distinct signatures of each phase as well, viz.,  
\begin{enumerate}
\item The BP1 phase is marked by two frequency branches ($\omega_1$ and $\omega_2$) of $\tilde{b}$ on each side of the resonance. i.e., there are a pair of solutions for each $\epsilon_b$. 
\item The BP2 phase is also marked by the pair of solutions for each $\epsilon_b$ as in BP1, but there is an additional solution $\omega_3$, too : resulting in three frequency branches on each side of the resonance. 
\item For the FFLO phase, the frequency spectrum is similar to that of BP1, only the $\omega$ values are slightly shifted. 
\item In the phase-separated state, there are two branches ($\omega_1$ and $\omega_2$) in the negative detuning side, and a single branch ($\omega_1$) on the positive detuning side. 
\end{enumerate}

Next we investigate the existence of these branches by analysing the mathematical structures of the relevant dynamical equations, and try to explain all the features mentioned above. It complements the numerical results shown in Sec. III. 

\subsubsection{The dominant branch : $\omega\approx -\epsilon_b$} 
In a non-interacting  system  with $g_1=0$ and $g_2=0$, Eq. (\ref{o2}) and (\ref{phi2}) take the simple form : 
\begin{equation}
\label{ononint}
i\hbar\dfrac{\partial \tilde{O}_{\textbf{p}}}{\partial t}=2\epsilon_\textbf{p} \tilde{O}_{\textbf{p}}
\end{equation}

\begin{equation}
\label{phinonint}
i\hbar\dfrac{\partial \tilde{b}}{\partial t}= \epsilon_b \tilde{b}
\end{equation}

Here we considered $\textbf{q}=0$ pairing only, but it can be extended to cover $\textbf{q}\neq 0$ FFLO pairing as well. 
So $\tilde{b}(t)\sim b_1 \mbox{e}^{-i\epsilon_b t/\hbar}$ and $\tilde{O}_p(t)\sim \tilde{O}_{p0} \mbox{e}^{-i 2\epsilon_p t/\hbar}$ 
Thus, for a completely non-interacting system, the bosonic field $b$ behaves as $$b(t)=b_0+\tilde{b}_1 \mbox{e}^{-i\epsilon_b t/\hbar}$$ and oscillates with a single frequency $\epsilon_b$. For a weekly interacting system as well, there will always be a frequency $\omega\sim\epsilon_b$, but there can be additional frequencies arising out of the coupling terms, as discussed next. 

\subsubsection{Emergence of A pair of frequencies}

We first consider the dynamics of the BP1 case. If $\omega >0$, Eq.(\ref{u1}) reduces to 
\begin{equation} 
\begin{split}
u(p,\omega)& =  \dfrac{1}{2\pi^2}\int \dfrac{p^2}{(p^2/m)+\hbar \omega}dp\\
=&\dfrac{1}{2\pi^2}\Big(mp-m\sqrt{m\hbar\omega} \tan^{-1}\Big(\dfrac{p}{\sqrt{m\hbar \omega}}\Big)\Big)\\
&\approx \dfrac{1}{2\pi^2}\dfrac{p^3}{3\hbar\omega}
\end{split}
\end{equation}
If, on the other hand,  $\omega <0$, we get
\begin{equation} 
\begin{split}
u(p,\omega)
=& \dfrac{1}{2\pi^2}\Big(mp-m\sqrt{m\hbar|\omega|} \tanh^{-1}\Big(\dfrac{p}{\sqrt{m\hbar |\omega|}}\Big)\Big)\\
\approx& \dfrac{1}{2\pi^2}\dfrac{p^3}{3\hbar\omega}
\end{split}
\end{equation}
These are obtained by expanding the $\tan^{-1}(p/\sqrt{m\hbar\omega})$ and $\tanh^{-1}(p/\sqrt{m\hbar|\omega|})$ functions, assuming $\hbar|\omega| >>  p^2/m$. This is valid in almost all our relevant parameter ranges as in our convention $\hbar=1$, $m=0.5$, $p$ can run from 0 to 1; while our plotted regions correspond to $\omega$ up to 40. 
\begin{equation}
u(p_3,\omega)-u(p_2,\omega) =\dfrac{1}{2\pi^2}\dfrac{p_3^3-p_2^3}{3\hbar\omega}=\dfrac{\alpha_1}{\hbar\omega}
\end{equation}
Where $\alpha_1 =(p_3^3-p_2^3)/6\pi^2$.\\

Putting this in Eq. (\ref{omega}), 
\begin{equation}
\hbar\omega+\epsilon_b+\dfrac{g_2^2\dfrac{\alpha_1}{\hbar\omega}}{1-g_1\dfrac{\alpha_1}{\hbar\omega}}=0\\
\end{equation}
If $g_1$ is negligibly small (which indeed is the case for the parameters of $^{40} \mbox{K}$ and $^{6} \mbox{Li}$), it reduces to 
\begin{equation}
(\hbar\omega)^2+\epsilon_b\hbar\omega -g_2^2\alpha_1=0
\label{omega1}
\end{equation}

This is a quadratic equation in $\omega$. So for each value of $\epsilon_b$, there will be two solutions for $\omega$:
\begin{equation}
\hbar\omega_{1,2}=\dfrac{1}{2}[-\epsilon_b \pm \sqrt{\epsilon_b^2-4g_2^2\alpha_1}]
\label{quad}
\end{equation}

If $\epsilon_b^2 < 4 g_2^2 \alpha_1$, there are no real solutions for $\omega$. This is why in the region near $\epsilon_b=0$ (on both sides of the resonance) there are no solutions at all. In the region $\epsilon_b >4g_2^2\alpha_1$, there are two solutions $\omega_1$ and $\omega_2$. When $\epsilon_b >>4\alpha_1$, $\omega_1= -\epsilon_b$ and $\omega_2=0$: so effectively, it is only one non-zero solution there.  This is why only a single $\omega$ branch survives in the large $|\epsilon_b|$ limit. 

So now in addition to the $\omega_1= -\epsilon_b$ solution as in the non-interacting system, an additional $\omega_2$ appears due to the presence of the interaction terms. For a large enough detuning value, $\omega_1$ would denote a high-frequency oscillatory component, and $\omega_2$ a low frequency oscillatory component in the fluctuation dynamics. If we considered more complicated states like BP2, FFLO or phase separation, the nature of the solutions would be the same; only the structure of $\alpha_1$ in Eq. (\ref{quad}) would change. Thus, $\omega_1$ and $\omega_2$ contain crucial information about (i) the coupling $g_2$ (and also $g_1$, if $g_1$ is non-negligible), (ii) the amount of imbalance present , and (iii) the nature of pairing (whether it is BP1, BP2, FFLO or phase separation).  It is the span of the breached region introduces an additional momentum scale in the system (for BP1 as discussed above, it is given by ($p_3-p_2$)) and that gets reflected in the frequency space being studied. 

\subsubsection{Appearance of the 3rd branch in BP2 }

So far, we have discussed solutions for $\omega$ under the assumption $\hbar|\omega|>> p^2/m$ only. Next we address solutions when $\omega$ is small, i.e., $\hbar|\omega|\sim p^2/m$. The function$\frac{p^2}{(p^2/m)+\hbar\omega}$ has a singularity if $\hbar\omega=-p^2/m$. As $\hbar$ is set to 1 and $m$ is taken as $0.5$,  for any $\omega$ in the range $0$ to $2$, there will be a $p$ (ranging from $p=0$ to $p=p_F=1$) for which $O_{\textbf{p}}$ diverges, and there are no solution for $\omega$ in this region.

In the BP2 structure, where there is a three-shell structure, and an intermediate region of unpaired fermions, this singularity can be bypassed if that $p$ is excluded from the sum just because it corresponds to the unpaired region.  Then,

\begin{equation}
\epsilon_b+ \hbar\omega+ g_2\sum_{\textbf{p}\epsilon \bar{\textbf{p}}} O_{\textbf{p}}=0
\end{equation}

Here $\bar{\textbf{p}}$ denotes the region where the fermions are paired. 

This equation yields a solution for $\omega$, which is  different from the pair of solutions $\omega_1$ and $\omega_2$ discussed previously. This corresponds to the 3rd branch of $\omega$ as seen in BP2 case. However, its exientence depends on the span of the breached region : whether it can exclude the singularity or not. Also, since the singularity occurs for negative $\omega$ only, $\omega_3$ would be always negative. 

\subsubsection{Phase separation : a single branch for positive detuning and two branches for negative detuning } 

The function $u_4(p,r_1,\omega)$ in Eq. (\ref{ups1}) can be expressed as a combination of the functions $\mbox{Ei}(i p r_1 + r_1\sqrt{m\hbar\omega})$, $\mbox{Ei}(-i p r_1 + r_1\sqrt{m\hbar\omega})$, $\mbox{Ei}(i p r_1 - r_1\sqrt{m\hbar\omega})$ and $\mbox{Ei}(-i p r_1 - r_1\sqrt{m\hbar\omega})$. Here Ei denotes exponential integral functions defined as $$\mbox{Ei}(x)=\int_{-x}^\infty \dfrac{e^{-t}}{t}dt$$

If $\omega >0 $, the argument of the Ei functions are in general complex, consisting of both real and imaginary parts. In our relevant parameter regime, $p<<\sqrt{m\hbar\omega}$, so $\mbox{Im}(u_4(p,r_1,\omega))<<\mbox{Re}(u_4(p,r_1,\omega))$. Applying $\mbox{Ei}(-x+i\delta)=-Ei(x)\mp i \pi$, it can be argued that $u_4(p,r_1,\omega)$, integrated between any two $p$ values will give a zero. As a result, $f_4(p_1,p_2,r_1,\omega)$ in Eq. (\ref{omegaps}) vanishes, and there is only one solution : $\hbar\omega_1 = -\epsilon_b$. 

If, on the other hand, $\omega <0$, the argument of the Ei functions are purely imaginary. It can be shown that in this case, $u_4(p,r_1,\omega)$, and hence  $f_4(p_1,p_2,r_1,\omega)$ is a real number, leading to  two solutions ( $\omega_1$ and $\omega_2$)  of the equation $\epsilon_b+\hbar\omega+g_2^2 f_4(p_1,p_2,r_1,\omega)=0$. Therefore, the $\omega <0$ solutions appear for both $\epsilon_b >0$ and $\epsilon_b<0$, while the $\omega >0$ solution appears only for $\epsilon_b <0$ as evident from Fig. \ref{plotps}.

\subsection{Simulation of actual dynamics : extraction of Oscillation frequencies}

Here we solve for the dynamics of the condensate fraction directly from Eq. (\ref{o2}) and Eq.(\ref{phi2}). Then going to the Fourier space, we extract the oscillation frequencies. The aim is to make a direct correspondance with the frequecies obtained from in Sec. III. We take the particular example of a BP1 state.  However, the dynamics of $\tilde{b}$ can be studied in the same fashion for the other pairing structures as well. 

We split both $\tilde{b}$ and $\tilde{O_p}$ in real and imaginary parts, and write $\tilde{b}=B_1+iB_2$ and $\sum_p\tilde{O_p}=A_1+iA_2$. We make an additional assumption that $g_1$ is negligibly small, an assumption that holds true in the systems that we considered. The paired region spans from $p_2$ to $p_3$ in the momentum space, as before. Equating the real and imaginary parts separately in Eq. (\ref{o2}), as well as in (\ref{phi2}), we obtain four coupled equations :

\begin{equation}
i\hbar \dfrac{\partial A_1}{\partial t}=2 A_2\big(\dfrac{p_3^3}{3}-\dfrac{p_2^3}{3}\big)+g_2B_2(p_3-p_2)
\end{equation}

\begin{equation}
i\hbar \dfrac{\partial A_2}{\partial t}=-2 A_1\big(\dfrac{p_3^3}{3}-\dfrac{p_2^3}{3}\big)-g_2B_1(p_3-p_2)
\end{equation}

\begin{equation}
i\hbar \dfrac{\partial B_1}{\partial t}=\epsilon_bB_2+g_2A_2
\end{equation}

and

\begin{equation}
i\hbar \dfrac{\partial B_2}{\partial t}=-\epsilon_bB_1-g_2A_1
\end{equation}

\begin{figure}
\includegraphics[scale=0.33]{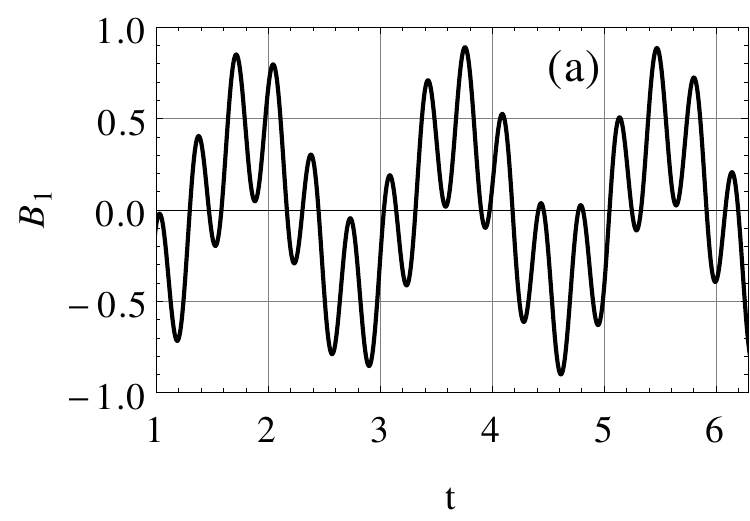}
\includegraphics[scale=0.33]{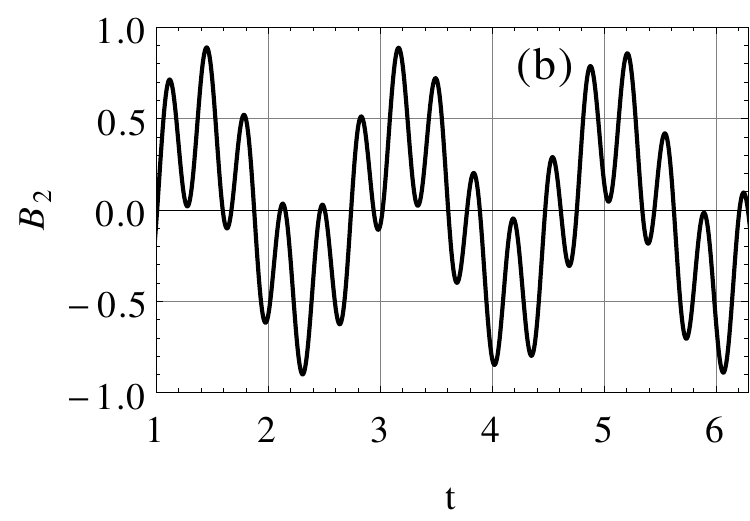}
\includegraphics[scale=0.33]{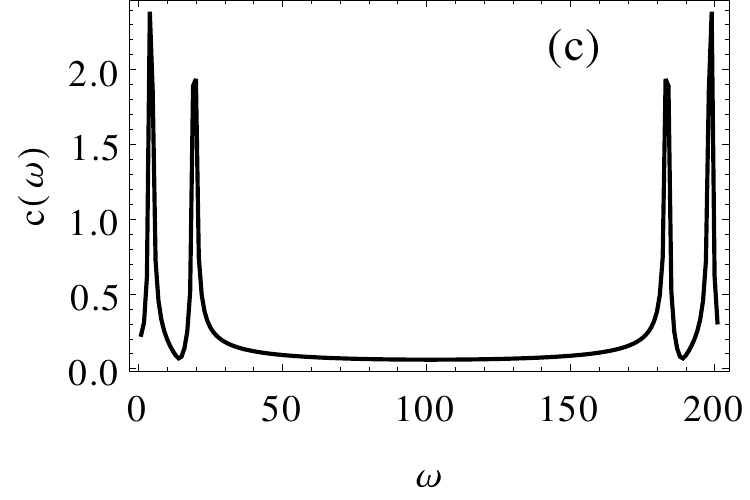}
\includegraphics[scale=0.33]{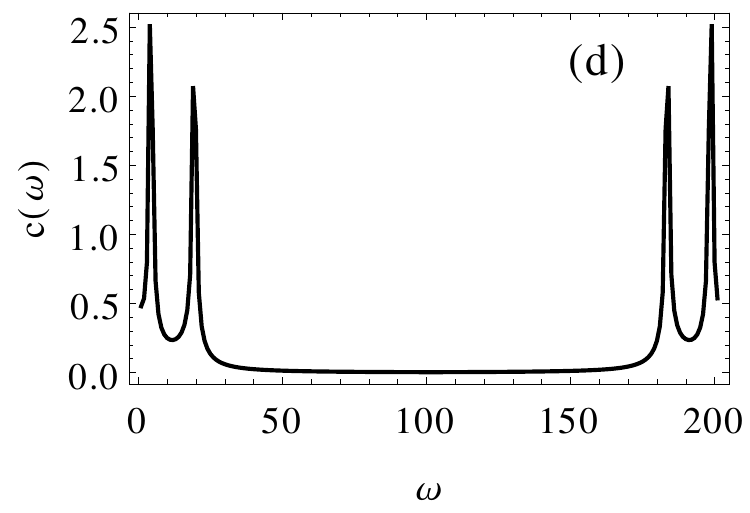}
\caption{Time evolution of $\tilde{b}(t)$ for $^{6}\mbox{Li}$, at $P=0.3$ and $\epsilon_b=-15$. Panels (a) and (b) : Time evolution of  Re[$\tilde{b}(t)$] and  Im[$\tilde{b}(t)$] respictively. Panels (c) and (d) :  Frequencies extracted from discrete Fourier transform ($N=200$) of Re[$\tilde{b}(t)$] vs. $t$ data and Im [$\tilde{b}(t)$] vs. $t$ data respectively. }. 
\label{fulldyn1}
\end{figure}

\begin{figure}
\includegraphics[scale=0.33]{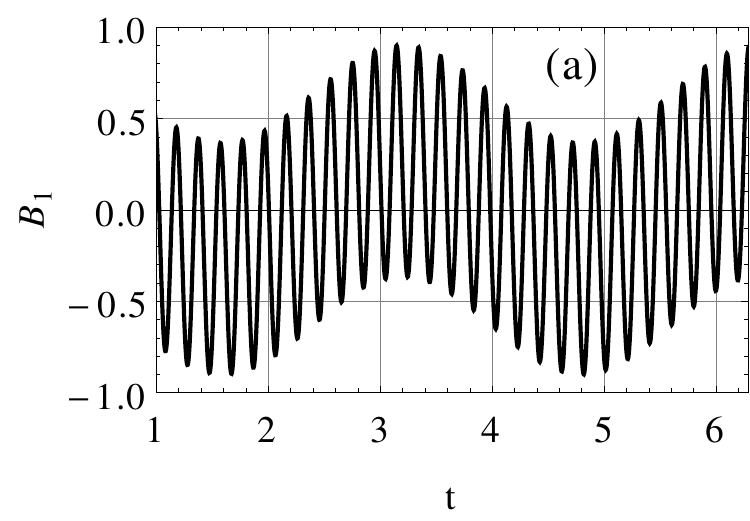}
\includegraphics[scale=0.33]{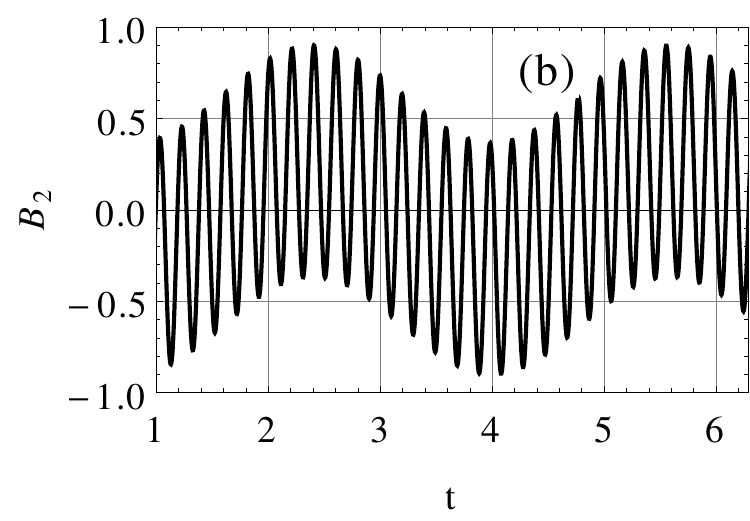}
\includegraphics[scale=0.33]{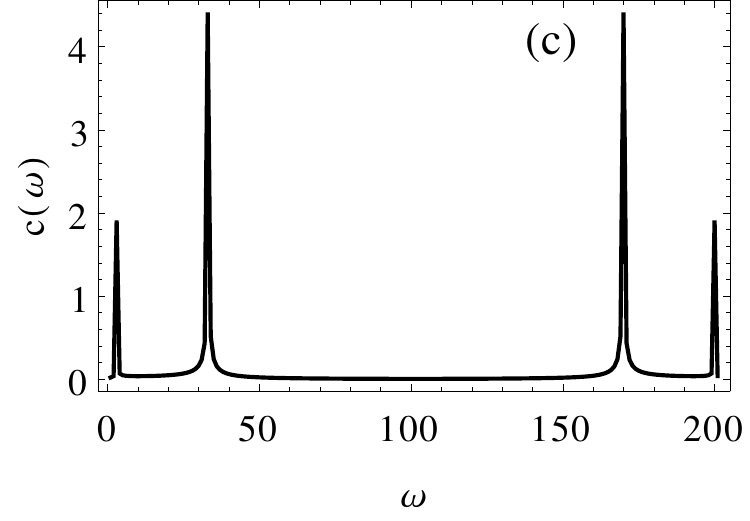}
\includegraphics[scale=0.33]{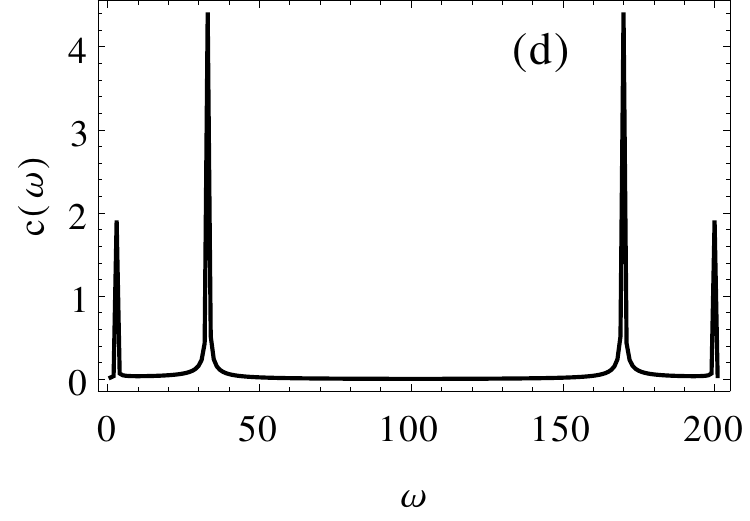}
\caption{Time evolution of $\tilde{b}(t)$ for $^{6}\mbox{Li}$, at $P=0.3$ and $\epsilon_b=-30$. Panels (a) and (b) : Time evolution of  Re[$\tilde{b}(t)$] and  Im[$\tilde{b}(t)$] respictively. Panels (c) and (d) :  Frequencies extracted from discrete Fourier transform ($N=200$) of Re[$\tilde{b}(t)$] vs. $t$ data and Im [$\tilde{b}(t)$] vs. $t$ data respectively. }
\label{fulldyn2}
\end{figure}

We solve these equations simultaneously. The time-evolution of  $B_1(t)$ (real part of $\tilde{b}(t)$) and $B_2(t)$(imaginary part of $\tilde{b}(t)$)are then plotted. From a list of 200 elements from the $\tilde{b}(t)$ vs. $t$ data, we take the discrete Fourier transform and extract the frequencies of oscillation. Here we consider a system of  $^{6}\mbox{Li}$ with a population imbalance of $P=0.3$. In Fig. \ref{fulldyn1},  we show the time evolution of $B_1(t)$, $B_2(t)$, and also their respective discrete Fourier transforms corresponding to $\epsilon_b=-15$. There appears two peaks at $\omega_1=\approx 18$ and $\omega_2\approx 5$;  and two symmetric peaks at $\approx (200-18)$ and $(200-5)$, indicating frequency components at $15$ and $5$ respectively. In Fig \ref{fulldyn2}, the same is repeated for $\epsilon_b=-30$. In this case, the frequencies obtained are $\omega_1\approx 30$ and $\omega_2\approx 0-1$ . These nearly match with the frequency spectrum shown in Fig. \ref{sfm}. Moreover, as evident from Fig. \ref{fulldyn1} and \ref{fulldyn2}, the frequencies are relatively closer when $|\epsilon_b|$ is small, but the separation increases with an increasing $|\epsilon_b|$; so the overall trend remains the same. 

\section{PROBING BY AN OSCILLATORY DRIVE}
As discussed in the previous section, the nature of the pairing can be obtained from the oscillation frequencies. These frequencies comfortably fall in the detectable range for ultracold atom experiments (for example, for BP1 in  $^{40} \mbox{K}$, $P=0.3$, the highest frequency shown in the plot is 30 in our unit with $\hbar=1$ and $E_F=1$;  that amounts to an actual frequency of $\sim 3.7$ MHz). However, a direct and exact measurement of the frequencies might turn out to be a little challenging. We describe below how the frequency can be determined accurately using an oscillatory magnetic field, via a method of resonance. 

To illustrate the scheme, we take the particular example of a two-shell BP1 structure, though it is applicable to BP2, FFLO and phase-separated states as well.

Let us add a small oscillatory component to the Feshbach magnetic field $H$, so that it
becomes $H(1 + \epsilon e^{i\Omega t} )$, with $\epsilon << 1$ . This is equivalent to replacing the factor $\epsilon_b$ by  $\epsilon_b (1 + \epsilon e^{i\Omega t} )$. We can make  perturbative expansions: $\Tilde {b}(t)=\Tilde{b^0}(t)+\epsilon \Tilde{b'}(t)$ and $\Tilde {O}(t)=  \Tilde{O^0}(t)+\epsilon \Tilde{O'}(t)$. Here $\Tilde{b^0}(t)$ and $\Tilde{ O^0}(t)$ are the values of $\tilde b(t)$ and $\tilde O(t)$ when there is no oscillatory part in the coupling. Noting that  $\Tilde{b_0}(t) = b_1 e^{i\omega_1 t}+b_2 e^{i\omega_2 t}$, it follows that
\begin{equation}
i\hbar \dfrac{\partial \Tilde{O'_p}}{\partial t}=2\epsilon_p \Tilde{O'_p}-g_1\sum_p O'_p-g_2 \Tilde{b'}
\end{equation}

\begin{equation}
i\hbar \dfrac{\partial \Tilde{b'}}{\partial t}=g_2\sum_p \Tilde{O'_p}+ \epsilon_b \Tilde{b'}+\epsilon_b e^{i\Omega t}(b_1 e^{i\omega_1 t}+b_2 e^{i\omega_2 t})
\end{equation}

 Taking Fourier Transforms as before, we find that $\Tilde{b'}(\omega)$ is non-zero only when $\omega=\omega_1+\Omega$ or $\omega=\omega_2+\Omega$. Its values at those two particular frequencies are 

\begin{equation}
\label{omega_match1}
\Tilde{b'}(\omega)=\dfrac{-\epsilon_b 2 \pi b_1}{\epsilon_b+\hbar(\Omega+\omega_1)+g_2^2f_1(p_2,p_3,\omega_1+\Omega))}\\
\end{equation}

and 

\begin{equation}
\label{omega_match2}
\Tilde{b'}(\omega)=\dfrac{-\epsilon_b 2 \pi b_2}{\epsilon_b+\hbar(\Omega+\omega_2)+g_2^2f_1(p_2,p_3,\omega_2+\Omega))}\\
\end{equation}

respectively. 

There is a resonance when the denominator becomes zero, i.e, $\epsilon_b+\hbar(\Omega+\omega_1)+g_2^2f(p_2,p_3,\omega_1+\Omega)=0$. But we know, $\epsilon_b+\hbar\omega_1+g_2^2f(p_2,p_3,\omega_1)=0$. Subtracting, 
\begin{equation}
\label{omega_find} 
\hbar\Omega+g_2^2(f(p_2,p_3,\omega_1+\Omega)-(f(p_2,p_3,\omega_1))=0
\end{equation}
and
\begin{equation}
\label{omega_find2} 
\hbar\Omega+g_2^2(f(p_2,p_3,\omega_2+\Omega)-(f(p_2,p_3,\omega_2))=0
\end{equation}
Since $\Omega$ is associated with the frequency of the time-dependent magnetic field, it is a tunable parameter. If, for a particular $\Omega$,  either of Eq. (\ref{omega_find}) or Eq.(\ref{omega_find2}) is satisfied, then we have a sharp resonance in the fluctuation of the condensate. It is obvious that $\Omega=0$ is a trivial solution of  Eq.(\ref{omega_find}) and Eq.(\ref{omega_find2}). To obtain non-zero solutions of $\Omega$, we solve it numerically. 

\begin{figure}[h]
\includegraphics[scale=.46]{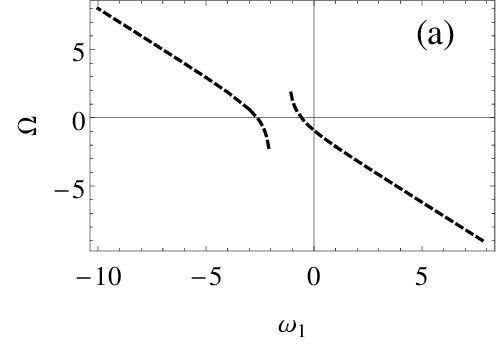}
\includegraphics[scale=.45]{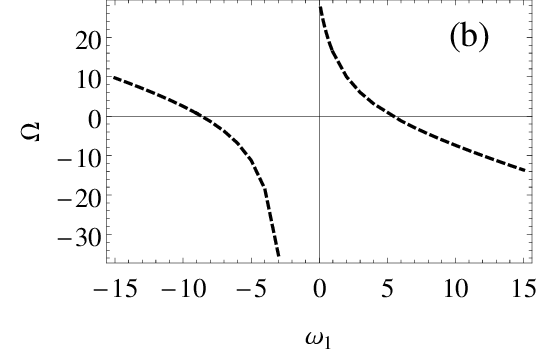}
\caption{$\Omega$  (frequency of the oscillatory field) vs. $\omega_1$ (natural frequency of oscillation of the condensate) for which there is a resonance in the condensate : (a) $^{40} \mbox{K}$, $P=0.3$ (b)  $^{6} \mbox{Li}$, $P=0.3$}
\label{omega2}
\end{figure}

In Fig. \ref{omega2}, the combinations ($\Omega$, $\omega_1$) are shown, for which there is a resonance. Fig. \ref{omega2}(a) corresponds to population-imbalanced $^{40} \mbox{K}$ with $P=0.3$. Fig. \ref{omega2}(b) is its counterpart for  $^{6} \mbox{Li}$.  So, if  $\Omega$ is gradually increased, keeping the system fixed in a particular detuning value, there will be two resonances: once when Eq. (\ref{omega_find}) is satisfied, and once when Eq. (\ref{omega_find2}) is satisfied. Since the value of $\Omega$ is already known, the value of $\omega_1$ can be extracted from the resonance conditions. Similarly, $\omega_2$ can be extracted, too.

 For the BP2 phase in an appropriate detuning regime, there will be three such $\omega$ values that can be detected by varying $\Omega$.

Once the  oscillation frequencies are obtained,  those values can be used to map the exact momentum-space configuration. For example, in case of the  BP1 structure, one can choose  guess values of $p_1$ and $p_2$ ( $p_3$ is set to 1 in our scale) so that the numerically computed oscillation frequencies closely match with those obtained in the experiment. Then, an iterative numerical calculation involving $p_1$, $p_2$, $p_3$ and $\omega_1$, $\omega_2$ will lead to the precise value of the breaching point.

\section{SUMMARY AND DISCUSSION }
Here we have studied the fluctuation dynamics of a population-imbalanced fermionic system capable of making a BCS-BEC crossover. We have shown that the fluctuations in the condensate fraction comprise a specific number of periodic components. This suggests that these oscillations do not die out with time; but are an integral part of the mean-field description of the system. 

We have also shown that the emergence of these oscillatory components depend on the momentum-space structures . For example, the BP1 state  and the FFLO phase are marked by a maximum of three oscillation frequencies in the fluctuation of the condensate fraction. The BP2 state, on the other hand, is marked by  six periodic components in the condensate dynamics.  A phase-separated state is characterized by three frequencies in one detuning side, and only one frequency in the other detuning side. We have analytically investigated the origin of all the periodic components in the fluctuation dynamics. We find that the nature and span of the unpaired region in each of the exotic phases introduce newer momentum scales in the dynamics : that get translated into these  frequencies of oscillation. 

It is also observed that  the $\omega$ values in the fluctuation of $\tilde{b}$ for the FFLO phase is slightly shifted from the corresponding $\omega$ value of the BP1 structure. The amount of this shift is proportional to the FFLO momentum $\textbf{q}$. The $\omega$ vs. $\epsilon_b$ response of the system can thus be used to uncover the detailed momentum-space structure of an FFLO type of superfluid. 

We have shown that if there is an oscillatory component in the Feshbach magnetic field, one can achieve a sharp resonance in the condensate fraction by tuning the frequency of the external magnetic field. This method can be used to detect the natural frequencies of oscillation of the condensate, which, in turn, give information about the pairing structures. 

Thus, the entire scheme is proven to be an indirect method of determining the momentum-space configurations of the imbalanced Fermi system. Since a direct experimental probe is not available, such an indirect method might would  out to be an efficient handle to detect the novel phases. 

The treatment described in this paper can be extended to detect more complicated structures in the momentum space, too. For example, usually the spin-polarized system appears as a mixture of various phases in a trap, and at the junction of two such phases, one has to additionally match the boundary conditions to solve the dynamical equations in each phase. Another possible extension of this work would be to apply this model to fermions in optical lattices, as  phases like breached pair and FFLO are now being extensively studied for lattice fermions \cite{kop, parish, wang, kinnu,cichy, karmakar}. 

\section*{ACKNOWLEDGEMENT}

R.D. would like to acknowledge support from the Department of Science and Technology, Government of India in the form of an Inspire Faculty Award (Grant No. 04/2014/002342).

\end{document}